\documentclass[reprint,aps,prl,superscriptaddress,amsmath,amssymb,floatfix,footinbib,nolongbibliography]{revtex4-1}
\usepackage[T1]{fontenc}
\usepackage{graphicx}
\usepackage{xcolor}
\usepackage[table]{xcolor}
\usepackage{bm}
\usepackage{soul}
\usepackage[normalem]{ulem}
\usepackage[colorlinks, linkcolor= blue, citecolor = blue, urlcolor=blue]{hyperref}
\usepackage{comment}
\usepackage{physics}

\usepackage{float}
\usepackage{pifont}
\usepackage{wrapfig}
\usepackage[stable]{footmisc}
\usepackage{array}
\usepackage{gensymb}
\usepackage{multirow}
\usepackage{hhline}
\newcommand{\cmark}{\ding{51}}%
\newcommand{\xmark}{\ding{55}}%
\usepackage{chemformula}
\def\nn{\nonumber}
\def\bea{\begin{eqnarray}}
\def\eea{\end{eqnarray}}
\def\be{\begin{equation}}
\def\ee{\end{equation}}

\def\tc{\textcolor}

\def\bal{\begin{aligned}}
\def\eal{\end{aligned}}

\begin{document}

\title{Deterministic Switching in Altermagnets via Asymmetric Sublattice Spin Current} 

\author{Sayan Sarkar}
\email{sayans21@iitk.ac.in}
\author{Sunit Das}
\email{sunitd@iitk.ac.in}
\author{Amit Agarwal}
\email{amitag@iitk.ac.in}
\affiliation{Department of Physics, Indian Institute of Technology, Kanpur-208016, India}

\begin{abstract} 
We demonstrate a deterministic switching mechanism in collinear altermagnets driven by asymmetric sublattice spin currents. Unlike conventional antiferromagnets, where combined parity-time-reversal symmetry enforces purely staggered sublattice spin torques, altermagnets host symmetry-protected nonrelativistic spin splitting that produces unequal torques on the two sublattices. Using doped FeSb$_2$ as a representative $d$-wave altermagnet, our Landau--Lifshitz--Gilbert simulations show that these torques enable magnetic-field-free and deterministic 180$^\circ$ N\'eel vector reversal over picosecond timescale. The mechanism is generic to even-parity altermagnets and remains effective even in centrosymmetric, weak spin-orbit coupled systems, where the N\'eel spin-orbit torque mechanism fails. Our results establish an experimentally accessible mechanism for switching of altermagnetic order, opening pathways for realizing ultrafast, low-power  altermagnet spintronic devices.  
\end{abstract}

\maketitle

\tc{blue}{\it Introduction--}  The detection and manipulation of magnetic order, particularly magnetization switching, is the cornerstone of spintronic memory devices~\cite{Tsymbal_19}. The speed and efficiency of switching ultimately dictate the competitiveness of emerging storage technologies~\cite{stuart_science2008, Miron_nature2011, Liu_Science2012, Wang_JOPD2013, Kent_NN2015, Wong_nature2015}. Conventional memories rely on ferromagnetic (FM) magnetization reversal, which offers straightforward readout, but FMs suffer from stray fields and limited dynamical speed~\cite{Shao_IEEE2021}. Although antiferromagnets (AFMs), with their compensated magnetization, provide ultrafast dynamics and immunity to stray fields, they lack a convenient electrical detection channel~\cite{Zhang_prb2022,wang_prl2023,kimel_JOM2024}. Recently discovered altermagnets (AMs) combine the advantages of both ferromagnets and antiferromagnets~\cite{smejkal_prx2022, Song_NRM2025, Rathore_JAC2025, Tamang_PP2025}. They host compensated magnetic order, ensuring ultrafast, stray-field-free operation like AFMs, while their large nonrelativistic spin splitting enables robust electrical readout through anomalous Hall signals~\cite{Gonzalez_prl23, Attias_prb2024, Reichlova_NC2024, Sato_prl2024, Tschirner_APL2023,Yao_arXiv2025}. These properties position AMs as a promising platform for high-speed, energy-efficient spintronic memory devices.  

Despite these advantages, deterministic control of altermagnetic order remains an outstanding challenge~\cite{Duan_prl2025,Gu_prl2025,Chen_prl2025}. In altermagnets with broken local inversion symmetry, the sublattice staggered N\'eel spin–orbit torque (NSOT) has been utilized to switch the N\'eel vector~\cite{zelenzy_prl2014,zelenzy_prb2017,Manchon_JOP2017,watanabe_prb2018,Jaurdan_nc18,Chen_prl2025,Zhou_NC2025, Wadley_nature24}. However, NSOT vanishes in most even parity altermagnets that have local inversion symmetry~\cite{Chen_prl2025,Zhou_NC2025}. Other proposals, such as N\'eel spin currents, rely on special sublattice geometries~\cite{Shao_prl2023} or require an auxiliary magnetic field~\cite{Feng_sa24, Shao_25}, limiting their device applicability. More recently, magnetic octupole injection has been proposed as a route~\cite{Han_25, Lee_prb25}, though 
control of octupoles in experiments remains difficult~\cite{Ye_nc24, Kimura_nc16}. These limitations highlight the pressing need for a generic, field-free switching mechanism applicable across the broader family of altermagnets.

In this Letter, we present a general switching mechanism for even-parity altermagnets based on asymmetric sublattice spin currents (ASSC), which arise naturally from the direction-dependent spin splitting of altermagnetic Fermi surfaces. The resulting imbalance of spin currents between distinct sublattices generates cooperative field-like and damping-like torques that drive N\'eel vector reorientation even in centrosymmetric altermagnets, where conventional NSOT vanishes. We show that ASSC arises intrinsically in heavy metal–altermagnet heterostructures, enabled by the spin-split symmetry of $d$-wave altermagnets. Using realistic parameters for doped \ch{FeSb2}~\cite{Smejkal_pnas2021}, we simulate the N\'eel vector dynamics with the Landau–Lifshitz–Gilbert equation and demonstrate ultrafast, deterministic 180$^\circ$ switching within a few tens of picoseconds ($\sim 40$ps). Importantly, the mechanism extends beyond $d$-wave to $g$- and $i$-wave altermagnets, establishing ASSC as a symmetry-allowed and broadly applicable route to field-free electrical switching of altermagnetic order, paving the way for ultrafast, energy-efficient spintronic devices.
\begin{figure}
    \centering
    \includegraphics[width=1\linewidth]{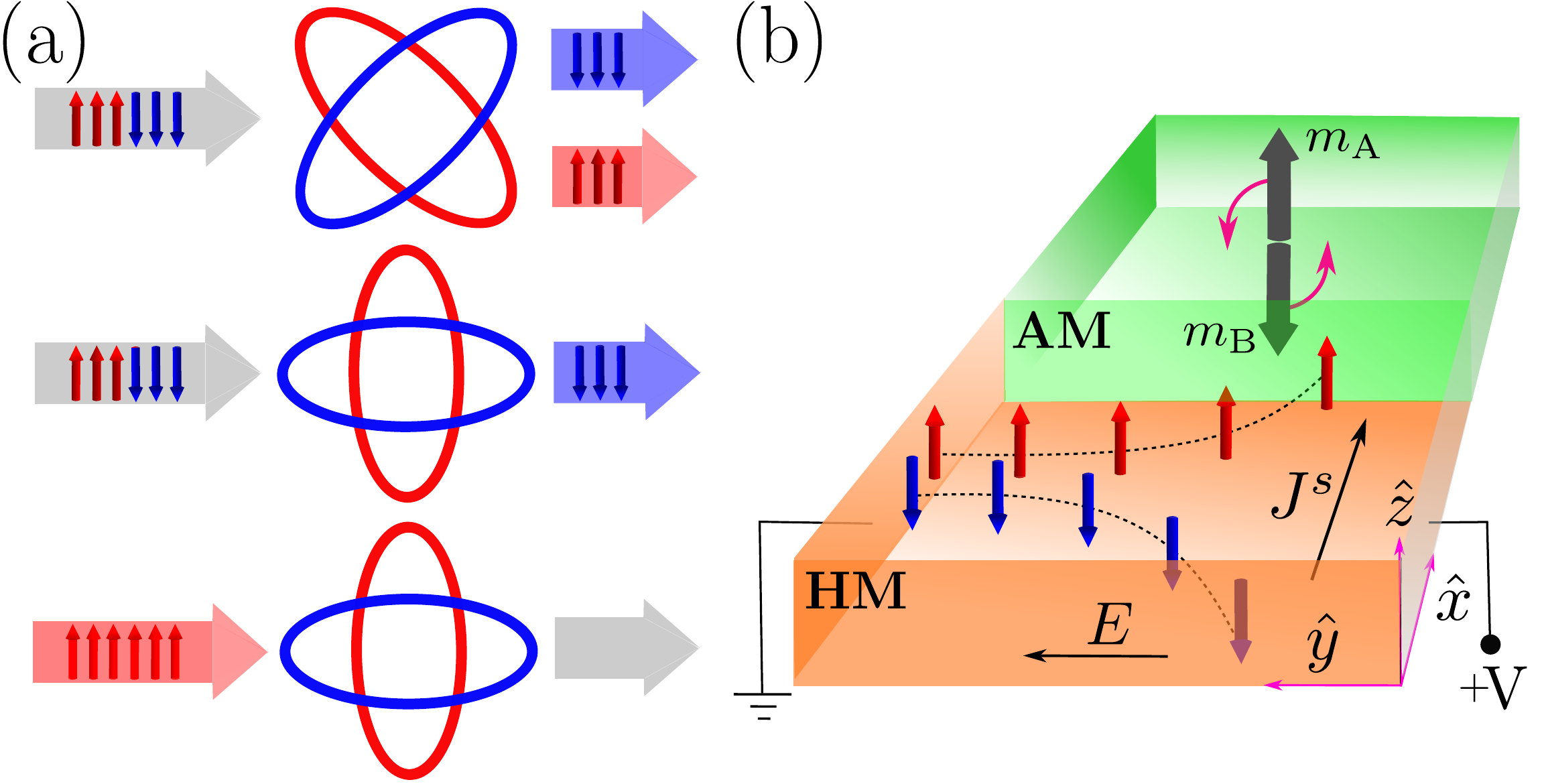}
    \caption{\textbf{Asymmetric sublattice spin current driven N\'eel switching in altermagnets.} (a) Direction-dependent spin conduction in an altermagnet. Conduction occurs when the injected spin polarization matches the spin character of the anisotropic Fermi surface. (b) Proposed altermagnet/heavy-metal (AM/HM) junction device, where the asymmetric sublattice spin current enables deterministic switching of the N\'eel vector.}
    \label{fig_1}
\end{figure}

\tc{blue}{\it ASSC in altermagnets--} 
Altermagnets belong to a distinct class of collinear magnets that host nonrelativistic spin-split band structures despite having vanishing net magnetization~\cite{smejkal_prx2022,Song_NRM2025}. Unlike conventional AFMs, where inversion-partner sublattices enforce equal and opposite ({\it i.e.,} staggered) spin torques, collinear altermagnets allow unequal sublattice spin currents and torques due to asymmetric spin splitting of the Fermi surfaces. This is the core principle behind asymmetric sublattice spin currents, which drive the N\'eel switching in our case. The spin splitting in altermagnets originates from exchange interactions and is dictated by spin space group symmetries~\cite{Reimers_NC2024, Yang_NC2025,jungwirth_arxiv2024}. This gives rise to direction-dependent spin-polarized Fermi surfaces. Collinear altermagnets realize even-parity spin splittings of $d$-, $g$-, or $i$-wave type~\cite{smejkal_prx2022,Xiao_prx2024}, all of which are symmetric under momentum inversion $\bm k \to -\bm k$. Among these, the $d$-wave case is the most widespread and has been intensively studied~\cite{sinova_prb2025, Zhao_prb2025, Sun_prb2025, Yan_arxiv2025, Liu_prb2025}. In this Letter, we focus on $d$-wave altermagnets, though our ASSC-based N\'eel switching mechanism is also applicable to other even-parity classes. 

The hallmark of $d$-wave altermagnets is the spin-group symmetry $[\mathcal{C}^2]\mathcal{C}^4_z \tau$~\cite{Turek_prb2022,smejkal_prx2022,Bhowal_prx2024,jungwirth_arxiv2024}, which combines a two-fold spin rotation $\mathcal{C}^2$, a four-fold real-space rotation $\mathcal{C}^4_z$, and a lattice translation $\tau$. This symmetry enforces equal spin textures and spin splittings along orthogonal directions [Fig.~\ref{fig_1}(a)]. The resulting spin-split Fermi surfaces are tied to opposite sublattices. Consequently, when an electric field is applied, the induced spin currents become unequal on the two sublattices, with their magnitude dictated by crystallographic orientation and direction of the field. We refer to this phenomenon as the asymmetric sublattice spin current (ASSC). Physically, the ASSC can be viewed as an intrinsic spin-filter: an applied current injects spins that preferentially couple to one sublattice over the other, generating unequal torques that collectively drive the N\'eel vector reversal.

\begin{figure}[t]
    \centering
    \includegraphics[width=1\linewidth]{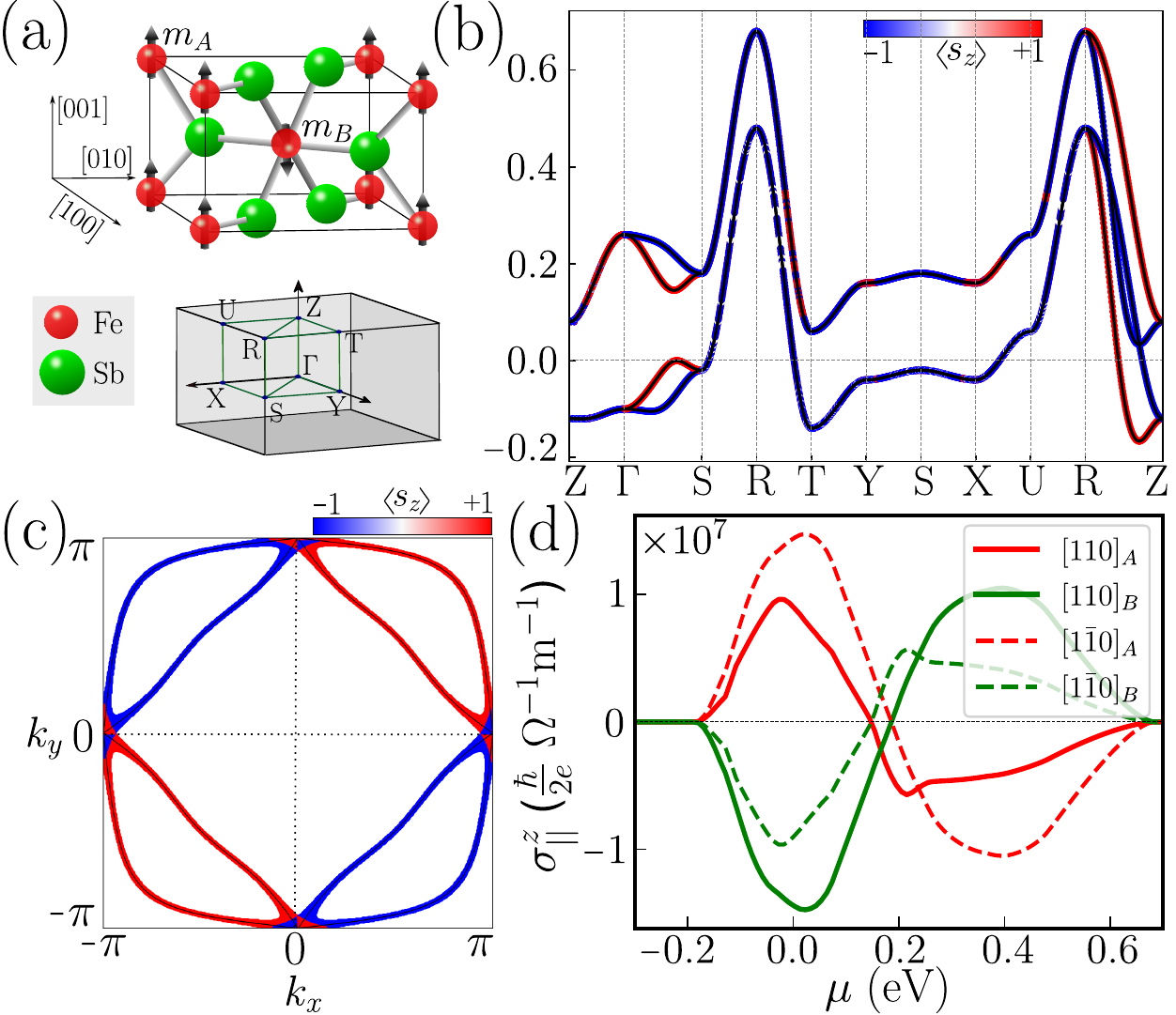}
    \caption{{\bf Origin of asymmetric sublattice spin current (ASSC) in doped $d$-wave altermagnet \ch{FeSb2}.} (a) Crystal structure and Brillouin zone. (b) Band structure of doped \ch{FeSb2} along $Z$-$\Gamma$-$\rm S$-$\rm R$-$\rm T$-$\rm Y$-$\rm S$-$\rm X$-$\rm U$-$\rm R$-$\rm Z$, showing nonrelativistic spin splitting along the $\Gamma$-$\rm S$ and $\rm Z$-$\rm R$ directions. (c) Spin-projected Fermi surfaces at $\mu=0.16~\rm eV$ in the $k_z=0$ plane, revealing the $d$-wave symmetry of the spin splitting. (d) Sublattice-resolved  longitudinal spin conductivities ($\hat z$-polarized) along $[110]$ and $[1\Bar{1}0]$ versus chemical potential. The unequal spin response on the $A/B$ sublattice along specific directions is a direct signature of ASSC.}
    \label{fig_3}
\end{figure}
\begin{figure*}[t]
    \centering
    \includegraphics[width=1\linewidth]{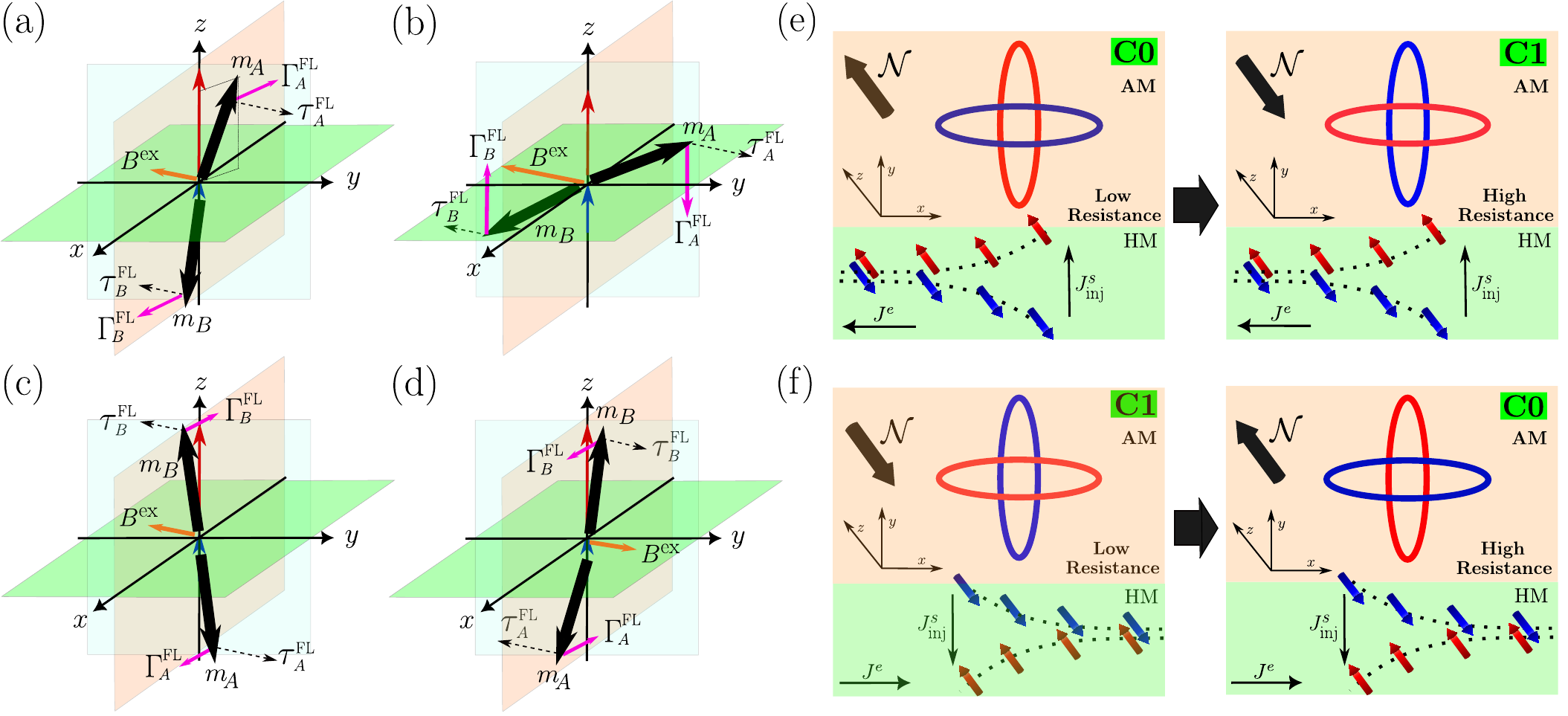}
    \caption{\textbf{Switching mechanism in an altermagnet/heavy-metal heterostructure device with a $d$-wave altermagnet.} (a)--(d) Torque profiles and effective field components governing the switching of the N\'eel vector by asymmetric sublattice spin current. Red and blue arrows indicate both the magnitude and polarization of spin currents acting on sublattices $A$ and $B$, respectively. (e) In the initial configuration (C0), the spin current polarization in the HM matches the anisotropic Fermi surface spin polarization of the AM along the current direction. This matching enables spin current transmission into the AM, generating ASSC-driven torques that switch the N\'eel vector ${\cal N} = ({\bm m}_A - {\bm m}_B)/2$ from C0 to C1. After switching, the Fermi surface polarization reverses, blocking further spin current transmission and stabilizing the high-resistance state. We verify this explicitly in Fig.~\ref{fig_4}. (f) The reverse transition (C1 $\rightarrow$ C0) is triggered by an oppositely polarized spin current, demonstrating deterministic control of the N\'eel vector via bias polarity reversal across the HM.}
    \label{fig_2}
\end{figure*}

Going beyond symmetry arguments, to demonstrate that ASSC is a measurable effect, we calculate the longitudinal spin conductivity $\sigma^{\nu}_{\parallel}$ of doped \ch{FeSb2}, a representative $d$-wave altermagnet~\cite{Smejkal_pnas2021,Shalika_arXiv2025}. 
While undoped \ch{FeSb2} is a Kondo insulator, electron-doped \ch{FeSb2} is a metallic $d$-wave altermagnet. We use a minimal tight-binding model consistent with its space group $Pnnm$~\cite{Roig_prb2024,Smejkal_pnas2021}, as described in EM1 of End Matter. The ${\bm k}$-dependent spin splitting originates from a term of the form $J_z \sigma_z \sin(k_x) \sin(k_y)$, where $\sigma_z$ denotes the spin operator and $J_z$ is the exchange parameter. Figure~\ref{fig_3}(a) and (b) show the crystal structure, Brillouin zone, and spin-polarized band dispersion.
The spin splitting is clearly visible along the $\Gamma$--$\rm S$ and $\rm R$--$\rm Z$ directions. Figure~\ref{fig_3}(c) illustrates the $d$-wave splitting of the spin-projected Fermi surfaces on the $k_z=0$ plane, showing identical spin character at $\bm k$ and $-\bm k$. We then evaluate the sublattice-resolved longitudinal spin conductivities of doped \ch{FeSb2} for $z$-polarized current along the $[110]$ and $[1\bar{1}0]$ crystallographic directions [Fig.~\ref{fig_3}(d)] (See EM2 of the End Matter for details of the calculation). Figure~\ref{fig_3}(d) clearly shows that the spin conductivities of the two sublattices, $\sigma^{\rm AM}_A$ and $\sigma^{\rm AM}_B$, differ in both $[110]$ and $[1\bar{1}0]$ directions. Additionally, the sign reversal between $\sigma^z_{\parallel A}$ and $\sigma^z_{\parallel B}$ along orthogonal directions is a direct fingerprint of the $d-$wave spin splitting of the Fermi surface, where the spin character changes sign under a $90^\circ$ rotation~\cite{Roig_prb2024}. Together, these features provide a clear picture of the origin of ASSC in $d$-wave altermagnet.  

We now focus on the emergence of ASSC in an AM/HM heterostructure [Fig.~\ref{fig_1}(b)] (see also EM3 of End Matter). An electric field in the HM layer generates a spin current via the spin Hall effect~\cite{Sinova_rmp2015,Hoffman_IEEE2013,Yang_prl2025,sarkar_arxiv2025}, which produces an effective electric field $E_{\rm int}$ on the AM through inverse SHE. This field can be expressed as the product of the HM longitudinal spin resistivity $\rho^{\rm HM}_{||}$ and the injected spin current $J^s_{\rm inj}$. The resulting sublattice spin currents in the AM are, 
\be
J^{s}_{A/B} = \sigma^{\rm AM}_{A/B} \,E_{\rm int} = \sigma^{\rm AM}_{A/B} \,\rho^{\rm HM}_{||}\, J^s_{\rm inj}~,
\ee
where $\sigma^{\rm AM}_{A/B}$ are sublattice-resolved spin conductivities of $A$ and $B$, and $J^{s}_{\rm inj}$ is the total injected spin current in the AM from the HM. Since $J^s_{\rm inj}$ and $\rho^{\rm HM}_{||}$ are fixed for a given AM/HM device and current direction, the asymmetry in the sublattice spin currents is dictated entirely by AM's spin conductivities along the current direction. Importantly, this ASSC in the heterostructure not only enables deterministic N\'eel vector switching, as we explore below, but also offers a symmetry-based control knob: the device orientation with respect to crystal axes can tune both the magnitude and sign of ASSC, thereby controlling the switching thresholds in AM/HM junctions. This symmetry-driven tunability of ASSC is a unique feature of altermagnets, absent in conventional AFMs and crucial for enabling deterministic spin control.

\textcolor{blue}{\it Magnetization dynamics under ASSC induced torques--} 
The dynamics of the N\'eel vector in an AM follows the modified Landau--Lifshitz--Gilbert (LLG) equations~\cite{slonczewski_JOM1996,Liu_prl2011,Yuan_IOP2020,Xu_JAP2023}. 
For a collinear AM with $A$ and $B$ sublattices, the evolution of the antiparallel sublattice magnetization unit vectors $\hat{m}_{A,B}$ is given by, 
\bea
     \frac{d \hat{m}_A}{dt} = -\gamma (\hat{m}_A \times {\bm B}^{\rm eff}_{A}) + \alpha (\hat{m}_A \times \frac{d \hat{m}_A}{dt}) + {\bm \tau}^{\rm FL}_{A} + {\bm \tau}^{\rm DL}_{A}, \label{Dynamic_AFM_A}~\\
     \frac{d \hat{m}_B}{dt} = -\gamma (\hat{m}_B \times {\bm B}^{\rm eff}_{B}) + \alpha (\hat{m}_B \times \frac{d \hat{m}_B}{dt}) + {\bm \tau}^{\rm FL}_{B} + {\bm \tau}^{\rm DL}_{B}.~ \label{Dynamic_AFM_B}
\eea
Here, $\gamma$ is the gyromagnetic ratio and $\alpha$ is the Gilbert damping coefficient. The effective field on each sublattice $i = A,B$ is ${\bm B}^{\rm eff}_i = {\bm B}^{\rm ans} + {\bm B}^{\rm ex}_i + {\bm B}^{\rm ext}$, combining the anisotropy, inter-sublattice exchange, and external magnetic field. The exchange field is ${\bm B}^{\rm ex}_{A/B} = -J_{\rm ex}\, {\hat m}_{B/A}$ with $J_{\rm ex}$ being the exchange coupling strength.  

The field-like (FL) and damping-like (DL) torques arise from the sublattice-resolved spin currents $J^{\nu a}_i$ in altermagnets, flowing along $a$ with spin polarization $\nu$. 
For a sublattice $i$, we have ${\bm \tau}^{\rm FL}_i = |\tau_i|\, \xi^{\rm FL}\,({\hat m}_i \times \hat{\sigma}^\nu)$, and $
{\bm \tau}^{\rm DL}_i = |\tau_i|\, \xi^{\rm DL}\, {\hat m}_i \times (\hat{m}_i \times \hat{\sigma}^\nu)$, with torque magnitude $|{\bm \tau}_i| = \frac{\gamma |J^{\nu a}_i|}{M_s t}$. Here, $M_s$ is the sublattice saturation magnetization, $t$ is the HM layer thickness, $\hat{\sigma}^\nu$ is the spin polarization direction of the injected current, and $\xi^{\rm FL/DL}$ are the torque efficiencies. $\xi^{\rm FL/DL}$ depends on spin-orbit coupling (SOC), spin-mixing conductance, interface orientation, and AM thickness~\cite{Sethu_pra2021,Amin_prb2016,Garello_nature2013,Manchon_RMP2019,Xu_JAP2023}. Solving these coupled equations yields the full dynamics of the sublattice magnetic moments and the N\'eel vector. 

\textcolor{blue}{\it Switching mechanism--} In presence of ASSC, polarized along the N\'eel vector direction [${\hat \sigma}^\nu={\hat z}$ for Fig.~\ref{fig_2}(a)], the magnetization dynamics is governed by the field-like torque ${\bm \tau}^{\rm FL}_i$ and the inter-sublattice exchange field ${\bm B}^{\rm ex}_i$. Small thermal fluctuations misalign  ${\hat m}_i$ from $\hat{ \sigma}^\nu$~\cite{Shao_prl2023}, activating ${\bm \tau}^{\rm FL}_i$ on both sublattices along with a precessional torque from ${\bm B}^{\rm ans}$. Since spin currents on $A$ and $B$ sublattices are unequal,  
the corresponding ${\bm \tau}^{\rm FL}_i$ are also unequal, forcing asymmetric tilts of ${\hat m}_A$ and ${\hat m}_B$, thereby inducing a finite canting. 
This produces an opposing exchange field ${\bm B}^{\rm ex}_i$. The resulting feedback introduces two effective torques, ${\bm \Gamma}^{\rm FL}_i = -\gamma \,({\hat m}_i \times {\bm B}^{\rm ex}_i)$ driving precession of the N\'eel vector around ${\bm B}^{\rm ex}_i$, and ${\bm \Gamma}^{\rm DL}_i = -\gamma \alpha\, {\hat m}_i \times ({\hat m}_i \times {\bm B}^{\rm ex}_i)$  resisting the canting (not drawn in Fig.~\ref{fig_2}). The interplay between ${\bm \tau}^{\rm FL}_i$ from the injected spin current and ${\bm \Gamma}^{\rm FL}_i$ from exchange field governs the dynamics illustrated in Fig.~\ref{fig_2}(a)-(d).

The switching proceeds as follows. As the N\'eel vector rotates toward the $xy$-plane due to ${\bm \Gamma}^{\rm FL}_i$, both ${\bm \tau}^{\rm FL}_i$ and ${\bm B}^{\rm ex}_i$ grow,  reaching a maximum when the N\'eel vector lies in the $xy$-plane [Fig.~\ref{fig_2}(b)]. Beyond this point they weaken, 
yet ${\hat m}_A$ and ${\hat m}_B$ continue to evolve until the N\'eel vector reaches the switched $180^\circ$ state [Fig.~\ref{fig_2}(c)].
If it overshoots, ${\bm \tau}^{\rm FL}_i$ reverses sign and restores the switched state [Fig.~\ref{fig_2}(d)]. This negative feedback stabilizes the final configuration. 

During reversal, the injected spin current is modulated by the orientation of the AM's spin-polarized Fermi surfaces, analogous to tunneling magnetoresistance. We capture this effect by introducing a phenomenological modulation factor,
\be \label{eta_fact}
   \eta(\theta) = \frac{1}{2}\left(1+\cos^2\frac{\theta}{2}\right),
\ee  
where $\theta$ is the angle between the N\'eel vector and the injected spin polarization. This reduces spin injection by up to 50\% when the N\'eel vector rotates by $180^\circ$. After switching, the Fermi-surface spin polarization also flips, blocking further injection and reinforcing stability. Reversal back to the initial state requires an oppositely polarized ASSC, achieved by reversing the HM bias [Figs.~\ref{fig_2}(e,f)].

The ASSC-driven protocol introduced here represents a qualitatively new mechanism for deterministic N\'eel vector switching and constitutes the central result of this Letter. As summarized in Table~\ref{Comparison}, previously proposed approaches, such as N\'eel spin–orbit torque (NSOT)~\cite{zelenzy_prl2014, zelenzy_prb2017, Chen_NM2019, Bodnar_NC2018}, N\'eel spin currents~\cite{Shao_prl2023, Shao_25, Fang_prb2024}, and magnetic octupole torques~\cite{Han_25, Lee_prb25}, either require broken local inversion symmetry, special sublattice geometries, or rely on multipole currents. NSOT, for example, produces purely staggered torques acting oppositely on inversion-partner sublattices~\cite{zelenzy_prl2014,zelenzy_prb2017,Bodnar_NC2018,Chen_NM2019}, and therefore vanishes in systems where each sublattice preserves local inversion symmetry. In contrast, ASSC operates in even-parity, centrosymmetric altermagnets without requiring relativistic spin–orbit coupling or external magnetic fields. 
Consequently, in materials like \ch{RuO2}, \ch{KV2Se2O}, and doped \ch{FeSb2}, where these conventional mechanisms may fail, the ASSC provides an experimentally accessible route to robust, field-free switching of the N\'eel vector.

Experimentally, N\'eel vector reversal can usually be detected through the accompanying sign change of the anomalous Hall voltage, providing a direct and robust electrical readout~\cite{Feng_nc22, Mathius_am25, Reichlova_nc24, Yu_npj25, gryglass_prb24, Maxim_prb24, Sajjan_prb25}. We show the sign reversal of anomalous Hall conductivity in doped FeSb$_2$ upon N\'eel switching in Fig.~\ref{fig_5}. 
While we focus on $d$-wave AMs, the same principle applies to $g$- and $i$-wave even-parity AMs, greatly expanding the material platforms for ultrafast low-power spintronic switching.

\begin{figure}[t]
    \centering
    \includegraphics[width=1.02\linewidth]{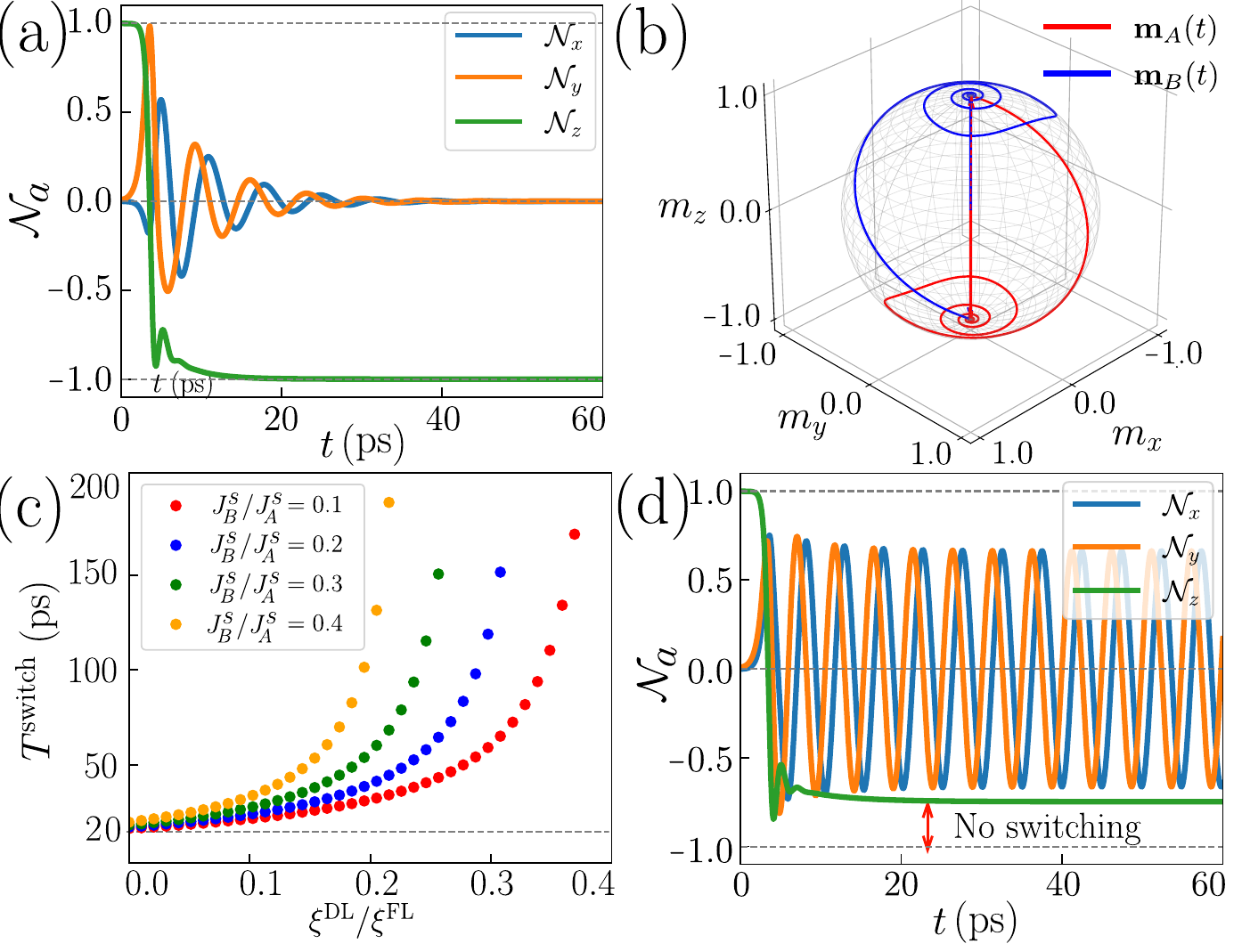}
    \caption{\textbf{Ultrafast N\'eel vector switching in the doped FeSb$_2$ device.} (a) Time evolution of the N\'eel vector components, showing full reversal of the $\hat z$ component within $40$ ps. (b) 3D trajectory of the sublattice magnetizations during the switching process. Together, (a) and (b) confirm robust and deterministic N\'eel vector reversal driven by ASSC-induced nonrelativistic torques. (c) Switching time of the N\'eel vector ($T^{\rm switch}$) as a function of torque efficiency ratio ($\xi^{\rm DL}/\xi^{\rm FL}$) for different asymmetric sublattice spin current ratios ($J^s_B/J^s_A$). (d) Persistent oscillation of the in-plane components of the N\'eel vector for a torque efficiency ratio of $\xi^{\rm DL}/\xi^{\rm FL}=0.5$ and $J^s_B/J^s_A=0.25$.}
    \label{fig_4}
\end{figure} 

\tc{blue}{\it N\'eel vector dynamics--} We validate the ASSC-based switching mechanism by solving the coupled LLG equations with realistic material parameters. Heavy metals such as Pt, Pd, Ta, and their alloys exhibit large spin Hall conductivities $\sim 10^5$–$10^6~(\hbar/2e)~\Omega^{-1}\rm m^{-1}$~\cite{kimura_prl2007, Ando_prl2008,Morota_prb2011,Liu_prl2011,Liu_Science2012,Zhu_pra2018,Zhu_SA2019}. For an electric field $E=10~\rm V/\mu m$, this generates a spin current $J^s\sim 10^{13}~(\hbar/2e)~\rm A/m^2$. Assuming a conservative $\sim 10\%$ transmission across the HM--AM interface~\footnote{Spin transmission across HM/FM or HM/AFM junctions is limited primarily by spin mixing conductance and spin memory loss. Both mechanisms reduce the effective spin current across the interface. Recent studies have proposed strategies to enhance spin transport efficiency~\cite{Tian_APL2024, Dong_APL2023, Zhu_prl2021, Biswajit_NL2025}. In our simulations, we adopt a conservative estimate of 10\% spin transmission loss, demonstrating that N\'eel vector switching via ASSC remains robust even under reduced spin transmission.}, the spin torque magnitude is $|{\bm \tau}^s|=\gamma J^s/M_s t\sim 1.54\times10^{11}~\rm s^{-1}$. This is equivalent to $B_{\rm eff}\sim 0.83~\rm T$, for sublattice magnetization of \ch{FeSb2}, $M_s\sim 1.5\times 10^{5}~\rm A/m$ (for individual Fe moment $\sim 1~\mu_B$~\cite{Smejkal_pnas2021,Lukoyanov_EPJ2006}) and HM thickness $t\sim 5~\rm nm$. We also assume an inter-sublattice exchange field $B^{\rm ex}\sim 190~\rm T$ and anisotropy field $B^{\rm ans}\approx 3.3~\rm T$~\cite{Smejkal_pnas2021}. For torque efficiencies, we take $\xi^{\rm DL}/\xi^{\rm FL}=0.25$ and the sublattice spin-current ratio $J^s_B/J^s_A=0.1$ (tunable via chemical potential). The angle-dependent modulation of ASSC torques [Eq.~\eqref{eta_fact}] is included throughout the simulations.

Using these parameters in Eqs.~\eqref{Dynamic_AFM_A} and \eqref{Dynamic_AFM_B}, we simulate the N\'eel vector dynamics for FeSb$_2$. Figure~\ref{fig_4}(a) shows the time evolution of its components. The $\hat z$-component reverses fully within $\sim 40$ ps, while $\hat x$ and $\hat y$ oscillate briefly before vanishing. The three-dimensional sublattice magnetization trajectories in Fig.~\ref{fig_4}(b) further confirm the coherent and deterministic nature of the switching pathway. Having established deterministic N\'eel vector switching, we now examine how the torque ratio $\xi^{\rm DL}/\xi^{\rm FL}$ and the sublattice spin-current asymmetry $J^s_B/J^s_A$ affect the dynamics. The field-like torque ${\bm \tau}^{\rm FL}$ drives switching by canting the antiparallel sublattices and initiating reversal, while the damping-like torque ${\bm \tau}^{\rm DL}$ counteracts the exchange-induced precessional torque ${\bm \Gamma}^{\rm FL}$ and slows the process. This competition is evident in Fig.~\ref{fig_4}(c): increasing $\xi^{\rm DL}/\xi^{\rm FL}$ at fixed ${\bm \tau}^{\rm FL}$ lengthens the switching time from $\sim 20$ ps to $\sim 200$ ps. In contrast, decreasing the ASSC ratio ($J^s_B/J^s_A$) accelerates reversal by strengthening the net field-like torque. However, when ${\bm \tau}^{\rm DL}$ becomes too large, deterministic switching fails and the N\'eel vector does sustained oscillations [Fig.~\ref{fig_4}(d)], reminiscent of a spin-torque oscillator regime~\cite{Cheng_prl16, Khymyn_nature17, Song_NC24}. Ultrafast deterministic switching is thus achieved when the field-like torque dominates, the damping-like torque is minimized, and the ASSC ratio is minimized. This qualitative behavior persists over a wide range of realistic material parameters, demonstrating that the mechanism is generic to $d$-wave altermagnets and not specific to FeSb$_2$. For typical Pt/FeSb$_2$ heterostructures, the estimated spin current density ($10^{12}-10^{13}~\hbar/2e~\rm A/m^2$) and switching time ($<40$ ps) are within experimentally accessible limits.

\tc{blue}{\it Conclusion--} To summarize, we have demonstrated that asymmetric sublattice spin currents (ASSC), intrinsic to altermagnets, provide a robust and deterministic route for N\'eel vector switching. Using realistic parameters for doped FeSb$_2$, we showed that unequal sublattice torques, generated via spin injection from a heavy metal, can drive ultrafast 180$^\circ$ reversal within tens of picoseconds. Unlike conventional spin-orbit torque mechanisms, this protocol relies on nonrelativistic, symmetry-protected spin splitting, making it broadly applicable to even-parity metallic altermagnets, including centrosymmetric systems with weak spin-orbit coupling. 

Our results establish a new switching mechanism and open a practical route to altermagnetic devices combining ultrafast dynamics with low-power operation. The crystalline orientation provides a direct means to tune asymmetric sublattice spin currents, enabling control over switching thresholds and current polarity, while the intrinsic anomalous Hall effect offers robust electrical readout of the N\'eel vector. Extending this approach to other symmetry classes of even-parity altermagnets and embedding it into nanoscale memory devices promises multifunctional spintronic platforms with high speed, efficiency, and symmetry-based tunability.

\tc{blue}{\it Acknowledgment---} We acknowledge many fruitful discussions with Nirmalya Jana (IIT Kanpur). S. S. acknowledges IIT Kanpur for funding support. S. D. is supported by the Prime Minister's Research Fellowship under the Ministry of Education, Government of India. A. A. acknowledges funding from the Core Research Grant by ANRF (Sanction No. CRG/2023/007003), Department of Science and Technology, India. 

\bibliography{refs}

\section{End Matter}

{\it \tc{blue}{EM1: Lattice model of $d$-wave altermagnet.}} 
We employ a minimal tight-binding model to capture the essential symmetry and electronic structure of $d$-wave altermagnets. We adapt the model parameters from Ref.~\cite{Roig_prb2024} for \ch{FeSb2} (space group $Pnnm$). It is a centrosymmetric antiferromagnet that exhibits $d$-wave spin splitting [see Fig.~\ref{fig_3}(c)]. The Hamiltonian is given by  
\be\label{Ham_AM}
{\cal H} = \varepsilon_{0,{\bm k}} + t_{x,{\bm k}} \tau_x + t_{z,{\bm k}} \tau_z + \tau_y \vec{\lambda}_{{\bm k}} \cdot \vec{\sigma} + \tau_z \vec{J} \cdot \vec{\sigma}~,
\ee
where $\tau_i$ and $\sigma_i$ are the Pauli matrices in orbital (sublattice) and spin space. Here, $\varepsilon_{0,{\bm k}}$ is the sublattice-independent dispersion, while $t_{x,{\bm k}}$ and $t_{z,{\bm k}}$ denote inter- and intra-sublattice hopping. $\vec{\lambda}_{{\bm k}}$ encodes spin-orbit coupling, and $\vec{J}$ is the N\'eel order parameter.
For the given space group, the coefficients are, 
%
\bea
    \varepsilon_{0,\bm k}&=&t_{1x}\cos{k_x}+t_{1y}\cos{k_y}+t_2\cos{k_z}+t_3\cos{k_x}\cos{k_y} \nn\\ && +
    t_{4x}\cos{k_x}\cos{k_z}+t_{4y}\cos{k_y}\cos{k_z}\nn\\&& +t_5\cos{k_x}\cos{k_y}\cos{k_z} - \mu \\
    t_{x,\bm k}&=& t_8\cos(k_x/2)\cos(k_y/2)\cos(k_z/2)\nn \\
    t_{z,\bm k} &=& t_6 \sin{k_x}\sin{k_y} + t_7\sin{k_x}\sin{k_y}\cos{k_z} \nn \\
    \lambda_{x, \bm k}&=&\lambda_{x0} \sin(k_x/2)\cos(k_y/2)\sin(k_z/2) \nn\\
    \lambda_{y, \bm k}&=&\lambda_{y0} \cos(k_x/2)\sin(k_y/2)\sin(k_z/2) \nn \\
    \lambda_{z, \bm k}&=& \lambda_{z0}\cos(k_x/2)\cos(k_y/2)\cos(k_z/2).\nn
\eea 
%
For \ch{FeSb2}, we use the following parameter values $t_{1x}=-0.1,t_{1y}=-0.05,t_2=-0.05, t_3=0.06, t_{4x}=0.1,t_{4y}=0.05,t_5=-0.05,t_6=0.05, t_7=-0.1, t_8=0.15$ and $J_z=0.1$ (all are in eV).

In the absence of SOC ($\lambda_{\bm k}=0$) with the N\'eel vector aligned along $\hat{z}$ direction ([001] axis), the dispersion simplifies to
\be
\varepsilon_{\pm} = \varepsilon_{0,{\bm k}} \pm \sqrt{t_{x, {\bm k}}^2 + (t_{z,{\bm k}} + J_z s_z)^2}~.
\ee
Here, $s_z=\pm1$ is the eigenvalue of $\sigma_z$ denoting the spin character of the band. This expression shows that the spin splitting arises from the cross term $t_{z,{\bm k}} \vec{J} \cdot \vec{\sigma}$, reflecting the interplay of intra-sublattice hopping and magnetic order. Since $t_{z,{\bm k}} \propto \sin{k_x} \sin{k_y}$, the spin splitting acquires a $d$-wave form factor, consistent with the symmetry constraints of the space group.

\begin{figure}
    \centering
    \includegraphics[width=0.8\linewidth]{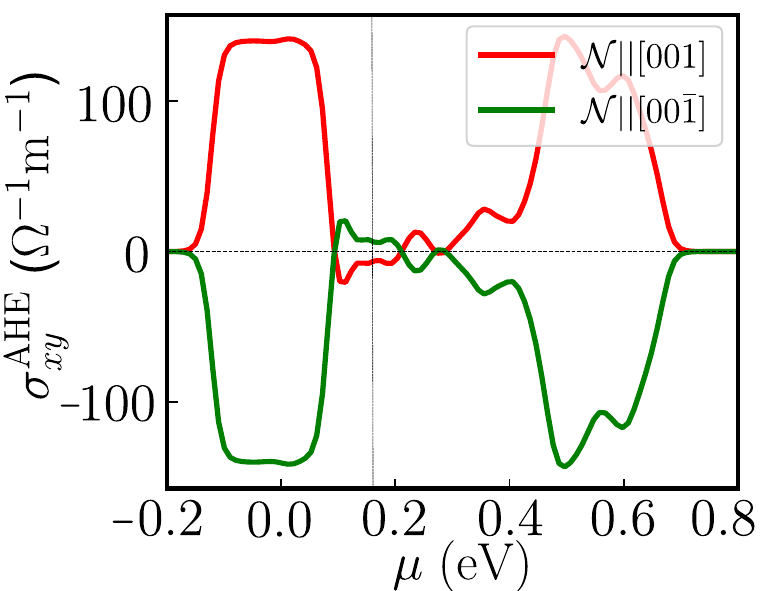} 
    \caption{{\bf N\'eel vector polarity dependent anomalous Hall effect.} Anomalous Hall conductivity of doped FeSb$_2$ as a function of chemical potential $\mu$ for the N\'eel vector oriented parallel (red curve) and antiparallel (green curve) to the [001] axis. The anomalous Hall response reverses sign across a broad range of $\mu$, providing a robust electrical signature of N\'eel vector reversal in the altermagnet. The vertical line marks the $\mu$ value, used for calculations in Fig.~\ref{fig_3}. 
    \label{fig_5}}
\end{figure}

With finite SOC, altermagnets can exhibit an anomalous Hall effect (AHE)~\cite{Gonzalez_prl23, Attias_prb2024, Reichlova_NC2024, Sato_prl2024, Tschirner_APL2023}. Assuming $\lambda_{x0}=\lambda_{y0} = 5~\mathrm{meV}$ and $\lambda_{z0} = 0$, we calculate the AHE conductivity as a function of chemical potential for opposite orientations of the N\'eel vector ($\mathcal{N}$), as shown in Fig.~\ref{fig_5}. The results clearly show that the AHE conductivity reverses sign upon switching the direction of the N\'eel vector. This provides a direct electrical readout of the N\'eel vector orientation. 
\begin{table*}[]
\centering
\small 
\renewcommand{\arraystretch}{1.1} 
\setlength{\tabcolsep}{4pt} 
\caption{{\bf Comparison of magnetic field-free switching mechanisms of N\'eel vector in antiferromagnets (AFMs) and altermagnets (AMs).} 
The proposed ASSC mechanism enables field-free, ultrafast switching in even-parity AMs without relying on relativistic SOC or staggered fields, distinguishing it from prior approaches. 
Abbreviations used: SOT -- spin–orbit torque, NSOT -- N\'eel spin–orbit torque, MOT -- magnetic octopole torque, AHE -- anomalous Hall effect, TMR -- tunneling magnetoresistance.}
\label{Comparison}
\begin{tabular}{cccccc} 
\hline \hline
\multirow{2}{*}{\textbf{Mechanism}} & \multirow{2}{*}{\textbf{Materials}} &  \textbf{SOC} & \multirow{2}{*}{\textbf{Readout}} & \multirow{2}{*}{\textbf{Key Features/Limitations}} & \multirow{2}{*}{{\bf Refs.}}\\
&& \textbf{required?} &  &  &  \\
\hline \hline
\multirow{3}{*}{NSOT} & AFMs and AMs & \multirow{3}{*}{\cmark} & \multirow{3}{*}{AHE} & Deterministic, SOC-driven switching; & \multirow{2}{*}{\cite{zelenzy_prl2014,zelenzy_prb2017,Bodnar_NC2018}}  \\
& with broken local &&& vanishes in centrosymmetric or even- & \multirow{2}{*}{\cite{Chen_NM2019,Chen_prl2025}}\\
& inversion symmetry &&& parity systems. & \\
\hline 
\multirow{2}{*}{Néel Spin} & \multirow{3}{*}{A and C type AFM} & \multirow{3}{*}{\cmark} & \multirow{3}{*}{TMR} & Field-free switching in selected AFMs;  & \multirow{3}{*}{\cite{Shao_prl2023,Fang_prb2024}}  \\ \multirow{2}{*}{Current} &&&& requires special sublattice geometry \\
&&&&  or weak magnetic field. & \\
\hline
\multirow{3}{*}{MOT} & \multirow{3}{*}{AM/HM heterostructure} & \multirow{3}{*}{\xmark} & \multirow{3}{*}{AHE} & Multipole-induced torque predicted; & \multirow{3}{*}{\cite{Han_25}} \\ &&&& {SOT is required along with MOT for} & \\
&&&& {deterministic switching.} & \\
\hline
\multirow{4}{*}{ASSC} & \multirow{3}{*}{AM/HM heterostructure} & \multirow{4}{*}{\xmark} &  \multirow{4}{*}{AHE} & \textcolor{black}{Deterministic switching of even parity} \\
 & \multirow{3}{*}{($d/g/i$-wave)} &&& metallic AMs; effective even in locally & {\bf This work} \\&&&& centrosymmetric AMs; tunable by crystal & \\
 &&&& orientation & \\
\hline \hline
\end{tabular}
\end{table*}

{\it \textcolor{blue}{EM2: Sublattice projected spin conductivity.}} 
The linear spin conductivity tensor ($\sigma^z_{a;b}$) is defined by  $J^{z}_a = \sigma^z_{a;b} E_b$, where $J^{z}_a$ denotes the spin current polarized along $z$ and flowing along the $a$ direction. The spin conductivity tensor has two distinct contributions: (i) a Drude-like spin conductivity and (ii) a spin–Berry curvature (SBC)–induced transverse conductivity~\cite{Sinitsyn_JPCM2008,Vignale_JOSM2009,Sinova_rmp2015}. The Drude term is $\mathcal{T}$-odd and vanishes in non-magnetic systems that preserve time-reversal symmetry ($\mathcal{T}$). In contrast, the SBC conductivity is $\mathcal{T}$-even and can appear in both magnetic and non-magnetic systems. In nonmagnetic heavy metals, the spin Hall current arises from SBC in the presence of spin–orbit coupling (SOC). However, in altermagnets with negligible relativistic SOC, the Drude-like  contribution is the dominant, and often the only, contribution to spin conductivity. The Drude-like spin conductivity is given by
\begin{equation}
\sigma^{z}_{a;b} = \frac{1}{2} \sum_m \int \frac{d^d k}{(2\pi)^d} 
\{ s^{z}, v^{a} \}_{mm} \frac{\partial f_0(\varepsilon_m)}{\partial k_b}~.
\end{equation}
Here, $d$ is the dimensionality, $m$ labels the bands, and $f_0(\varepsilon_m)$ is the Fermi–Dirac distribution for the $m$-th band. The spin current operator is $(1/2)\{s^z, v^a\}_{mm} = (1/2)\langle u_m | (s^z v^a + v^a s^z) | u_m \rangle$, with $|u_m\rangle$ being the eigenstate of the $m$-th band. 

To demonstrate the asymmetric spin conductivity between the two sublattices of AM in the AM/HM heterostructure [see Fig.~\ref{fig_2}], we calculate the sublattice-resolved spin conductivity along a specific crystallographic direction. This is obtained by projecting $\ket{u_m}$ onto the $A$ and $B$ sublattices and evaluating the sublattice-projected spin current operator, $(1/2)\{s^z,v^a\}_{mm}^{A/B}=(1/2)\bra{u_{m}}{\hat P}_{A/B}^{\dagger}(s^z v^a+v^a s^z){\hat P}_{A/B}\ket{u_m}$. Here, ${\hat P}_{A/B}$ is the sublattice projection operator onto the $A/B$ sublattice. For the minimal Hamiltonian of doped \ch{FeSb2} in Eq.~\eqref{Ham_AM}, the basis is  $\ket{A,\uparrow},\ket{A,\downarrow},\ket{B,\uparrow},\ket{B,\downarrow}$. The corresponding sublattice projection operators are, 
\be
P_A=\frac{1}{2}(\tau_z + \tau_0)\otimes \sigma_0
;~~~~P_B=
\frac{1}{2}(\tau_z-\tau_0)\otimes \sigma_0.
\ee
Here, $\tau_i$ and $\sigma_i$ are Pauli matrices in the sublattice and spin space, as  in Eq.~(\ref{Ham_AM}).

{\it \tc{blue}{EM3: Spin transport across AM/HM junction.}} 
Spin transmission across an altermagnet/heavy metal (AM/HM) interface has been theoretically investigated in several recent works~\cite{Das_JOP2023, Lyu_RP2024}, which reveal direction- and spin-dependent carrier injection through such junctions. These studies show that the transmission amplitude depends sensitively on multiple factors, including the spin polarization ($\hat{\sigma}_z$) of the electron, 
the hopping mismatch between HM and AM ($t/t'$), and the applied bias ($V$) across the junction. In particular, transmission is enhanced when the spin of the injected electron aligns with the spin texture of the AM along the dominant axes of its spin-split Fermi surface. 

In our AM/HM geometry, the spin Hall voltage ($V_{\text{sh}}$) in the HM effectively serves as the spin injection bias. A higher $V_{\text{sh}}$  increases the spin current across the interface, as explicitly demonstrated in Ref.~\cite{Das_JOP2023}. Owing to the collinear nature of $d$-wave altermagnets such as doped \ch{FeSb2}, spin-up and spin-down states reside predominantly on opposite sublattices. As a result, spin-polarized injection from the HM couples asymmetrically to the two magnetic sublattices of the AM. This asymmetry is evident in the sublattice-projected longitudinal spin conductivities shown in Fig.~\ref{fig_3}(d), where an imbalance between the $A$ and $B$ sublattices is observed across injection directions. The injected spin current flows through the sublattices in proportion to their respective conductivity magnitudes  ($|\sigma^s_{A}|/|\sigma^s_{B}|$). This provides the microscopic origin  of the asymmetric sublattice spin currents (ASSC) used in our model.

{\it \tc{blue}{EM4: Other mechanisms of N\'eel vector switching.}} 
Table~\ref{Comparison} summarizes and compares the recently proposed N\'eel vector switching mechanisms. In contrast to other approaches, the ASSC based mechanism proposed here originates from the intrinsic, nonrelativistic spin splitting of even-parity altermagnets, and does not require any spin–orbit torque, staggered exchange field, or symmetry breaking. Its magnitude and polarity can be tuned solely by the crystalline orientation and current direction, offering an experimentally accessible platform for field-free, ultrafast, and low-power altermagnetic switching.


{\it \tc{blue}{EM5: Other altermagnets.}} In the main text, we demonstrated ASSC-mediated switching in AM/HM heterostructures using the doped $d$-wave altermagnet \ch{FeSb2} as an explicit example. The same mechanism applies to other rutile-structure altermagnets, such as \ch{RuO2}$[110]$/HM, and \ch{KV2Se2O}$[100]$/HM. For more efficient and faster switching in even-parity collinear altermagnets, including $g$- and $i$-wave systems, two conditions should be optimized: i) the AM layer should be epitaxially aligned with the heavy metal so that the crystallographic axis with strongest spin splitting in $\bm{k}$-space coincides with the HM's spin Hall current direction; and ii) the injected spin current should be polarized parallel to the N\'eel vector. Candidate heterostructures for $g$-wave altermagnets satisfying these criteria include \ch{MnTe}$[11\bar{2}0]$/HM~\cite{Krempask_Nat2024,Takahashi_NPJ2025} and \ch{FeS2}$[120]$/HM~\cite{Sanyal_prb2021,Liu_prb2025} heterostructures.  

In some cases, such as \ch{MnTe}, the N\'eel vector lies in-plane. Here,  collinearly polarized spin Hall currents from the HM, rather than conventional transverse ones, may provide more efficient switching. Material platforms predicted to exhibit this collinear spin Hall effect have been proposed in Refs.~\cite{Yang_prl2025,sarkar_arxiv2025}, and  can be used to construct suitable heterostructures.

\end{document}